\documentclass[a4paper,11pt]{article}

\usepackage{jheppub} 

\usepackage{empheq,mathtools}
\usepackage{amsmath,amsfonts,amssymb}
\usepackage{mathrsfs}
\usepackage{cancel,slashed}
\usepackage{tensor}
\usepackage{xcolor}
\usepackage{graphicx}
\usepackage{subfigure}
\usepackage{float}

\makeatletter
\def\@fpheader{\relax}
\makeatother

\newcommand{\be}{\begin{equation}}
\newcommand{\ee}{\end{equation}}

\usepackage{ulem}

\usepackage[czech,english]{babel}
\usepackage{yfonts}

\usepackage{url}
\usepackage{mathabx}

\usepackage{tikz}
\usetikzlibrary{positioning, arrows.meta}

\title{\boldmath Supersymmetric Spectral Form Factor of ABJM Theory}

\author{Wenni Zheng\,}

\affiliation{International Centre for Theoretical Physics Asia-Pacific (ICTP-AP), University of Chinese Academy of Sciences, 100190 Beijing, China}

\emailAdd{zhengwenni22@mails.ucas.ac.cn}

\abstract{We investigate the large-$N$ index analog of the spectrum form factor for ABJM theory in the microcanonical ensemble. In the  Cardy-like limit, the most dominant saddle describing the dual black hole decays rapidly at early times. However, the late-time behavior of the spectral form factor is determined by multi-cut saddles, which prevent it from decaying.}

\begin{document} 
\maketitle
\flushbottom

\section{Introduction}\label{sec:intro}


The spectral form factor (SFF) serves as a powerful probe of the pair correlation function of energy eigenvalues and a robust diagnostic of quantum chaos~\cite{Papadodimas:2015xma, Cotler:2016fpe, Liu:2018hlr}. It exhibits a universal ``dip-ramp-plateau'' structure in chaotic systems: an initial decay is followed by a linear ramp, signifying spectral rigidity, and finally a saturation to a plateau. This time evolution reveals statistics at increasingly finer energy scales. As such,  the SFF does not vanish at late times in a typical quantum system with a discrete spectrum.

Originally introduced in the study of quantum chaos and later adapted to gravitational and holographic settings, the SFF provides a refined probe of the fine-grained energy spectrum. A profound challenge arises in the context of quantum gravity. The semiclassical geometric description is dominated by a black hole saddle point, which yields a rapidly decaying SFF at early times but fails to capture the required non-vanishing behavior at late times. Reconciling this discrepancy by finding a gravitational origin for the full ``dip-ramp-plateau" structure is thus a central pursuit.

In gravitational systems, the spectral form factor serves as a key observable for probing the discreteness of the microscopic spectrum. In the context of low-dimensional models such as JT gravity~\cite{Jackiw:1984je, Teitelboim:1983ux}, the characteristic ``ramp'' behavior of the SFF is understood within an ensemble-averaged framework, where contributions from wormholes in Euclidean spacetime, particularly the double-cone geometry~\cite{Saad:2018bqo, Mahajan:2021maz, Cotler:2021cqa}, become dominant at intermediate times. Nevertheless, this ensemble interpretation raises the well-known factorization puzzle in holography. A central challenge, therefore, is to identify gravitational configurations or mechanisms that can produce the ramp without invoking ensemble averaging, factorizing contributions such as single-boundary ``half-wormhole" saddles~\cite{Saad:2021rcu, Mukhametzhanov:2021nea, Peng:2022pfa} or other yet-unidentified semiclassical geometries that respect the underlying discreteness of the spectrum.

The superconformal index (SCI) of superconformal field theories, viewed as supersymmetric partitions on $S^1\times S^2$ from the UV descriptions, encodes the protected spectra of BPS local operators~\cite{Romelsberger:2005eg, Kinney:2005ej}. As shown in \cite{Cabo-Bizet:2018ehj, Choi:2018hmj, Benini:2018ywd}, the SCI of four-dimensional $\mathcal{N}=4$ super-Yang-Mills theory with complex chemical potentials successfully reproduces the Bekenstein-Hawking entropy of the dual AdS$_5$ rotating BPS black holes. Similar analyses for AdS$_4$ rotating black holes~\cite{Chong:2004na} in terms of ABJM SCI have been performed in~\cite{Choi:2019zpz, Choi:2019dfu, GonzalezLezcano:2022hcf, Bobev:2022wem, BenettiGenolini:2023rkq} (see also~\cite{Nian:2019pxj, David:2020ems, David:2020jhp, Nian:2025iei}).

The spectral form factor calculated in the canonical ensemble, conventionally defined as the squared modulus of the analytically continued partition function $|Z(\beta+it)|^2$~\cite{Papadodimas:2015xma},  i.e., the Fourier transform of the eigenvalue pair correlation function, exhibits a fundamental limitation: its decay at early times is insufficiently rapid to probe the fine-grained structure of the spectrum effectively. This limitation is decisively overcome by transitioning to a microcanonical framework~\cite{Gharibyan:2018jrp, Saad:2018bqo}. The microcanonical SFF, denoted $|Y_{E,\Delta}(t)|^2$,  is constructed by  performing a convolution of the spectral density with a Gaussian energy window of width $\Delta$ centered at $E$. The modulating weight $\exp[-(E_n-E)^2/(2\Delta^2)]$, on the one hand, isolates a specific high-energy sector, preventing the spurious dominance of low-energy configurations (like thermal AdS geometry) at later times, which is a common pitfall when studying $Z(\beta+it)$ for black holes. On the other hand, this modulating factor induces a sharply accelerated initial decay, typically Gaussian: $|Y_{E,\Delta}(t)|^2\sim e^{-\Delta^2t^2}$. This rapid decay starkly amplifies the ``early-time challenge'' to the discreteness of the spectrum, as the semi-classical prediction plummets to an exponentially small value on an $\mathcal{O}(1)$ time scale.

The utility of this microcanonical formulation is exemplified in studies of free large-$N$ Yang-Mills theories, where subdominant saddles at early times become dominant as time evolves, preventing the SFF from decaying to exponentially small values~\cite{Chen:2022hbi}. An index-like generalization of the microcanonical SFF was later introduced to probe the microscopic spectrum of four-dimensional $\mathcal{N}=4$ SYM theory~\cite{Choi:2022asl}. It was found that the multi-cut saddle points govern the late-time behavior of the SFF, and these correspond holographically to $\mathbb{Z}_K$-orbifolded Euclidean black holes in AdS$_5\times S^5$~\cite{Aharony:2021zkr}.

In this manuscript, we consider the SCI version of the microcanonical SFF for the ABJM theory, which is dual to AdS$_4$ rotating BPS black holes. We will show that the time evolution of the spectral form factor (SFF) progresses through distinct dynamical regimes. Following an initial transient rise, the SFF enters the characteristic early-time ``slope'' phase, marked by a rapid exponential decay. This decay is subsequently hindered as contributions from multi-cut saddles become dominant.

This paper is organized as follows.  In Sec.~\ref{sec:SFF}, we review some basic concepts of the spectral form factor. In Sec.~\ref{sec: SFF-ABJM}, we introduce multi-saddles into the superconformal index of the ABJM theory, and use the new index to study the spectral form factor. Finally, we conclude with future directions in Sec.~\ref{sec:discussion}.

\section{Spectral Form Factor}\label{sec:SFF}

In this section, we briefly review the definition and essential properties of the spectral form factor, which will serve as the central quantity in the subsequent analysis.

The canonical spectral form factor is defined as the square modulus of the analytically continued  partition function~\cite{Papadodimas:2015xma}:
 \begin{align}
    \big|Z(\beta,t)\big|^2=Z(\beta+it)Z(\beta-it)\,,
   \end{align}
   where
 \begin{align}
    Z(\beta+it)=\sum_ne^{-(\beta+it)E_n}\,.
  \end{align}
  As the time $t$ increases, the phase factors $e^{-iE_nt}$ cause increasingly rapid oscillations among high-energy states. This leads to a destructive interference that effectively lowers the effective temperature of the system. In holographic setups, this effect implies that the dominant gravitational saddle switches from the black hole geometry to the thermal AdS geometry around $t\sim\mathcal{O}(\beta)$, thereby obscuring the late-time behavior of the black hole microstates.

To isolate the high-energy sector and avoid this spurious dominance of low-energy configurations, a microcanonical version of the SFF was introduced~\cite{Gharibyan:2018jrp}. It is defined by inserting a Gaussian energy window of width $\Delta$ centered at energy $E$
\begin{align}
   Y_{E,\Delta}(t) = \sum_ne^{-\frac{(E_n-E)^2}{2\Delta^2}}e^{-iE_nt}\,.
\end{align}
For a system with a continous energy spectrum, the sum can be replaced by an integral over the density of states $\rho(E)$
\begin{align}
  Y_{E,\Delta}(t) = \int d\tilde{E}\, \rho(\tilde{E})\, e^{-\frac{(\tilde{E}-E)^2}{2\Delta^2}}e^{-i\tilde{E}t}\,,
\end{align}
where $\rho(E)$ is obtained from the partition function via an inverse Laplace transform
\begin{align}
  \rho(E) = \int_{\mathcal{C}} \frac{d\beta}{2\pi i}\, e^{\beta E}\, Z(\beta)\,.
\end{align}
Carrying out the Gaussian integral over $\tilde{E}$ yields a convenient integral representation that will be used in the following section
\begin{align}
  Y_{E,\Delta}(t) = & \int_\mathcal{C} d\beta\, \frac{\Delta}{\sqrt{2\pi}i}\, e^{(\beta-it)E+\frac{1}{2}(\beta-it)^2\Delta^2}\, Z(\beta)\nonumber\\
  = & \int_\mathcal{C} d\beta\, \frac{\Delta}{\sqrt{2\pi}i}\, e^{\beta E+\frac{1}{2}\beta^2\Delta^2}\, Z(\beta+it)\,.
\end{align}
This microcanonical formulation filters out the low-energy thermal AdS contributions and retains sensitivity only to states within an energy window of order $\Delta$ around $E$. As a result, it enables a clean study of the black-hole-like high-energy states even at late times, where the canonical SFF $|Z(\beta,t)|^2$ would otherwise be dominated by the low-energy saddle.

\section{Spectral Form Factor of ABJM theory}\label{sec: SFF-ABJM}

We now provide a concise review of the large-$N$ superconformal index (SCI) of the ABJM theory in the Cardy-like limit $\beta\to 0$, and extend the construction to include multi-cut saddle-point configurations. This framework will subsequently be used to evaluate the microcanonical spectral form factor for the theory.

The SCI of the ABJM theory is defined as a supersymmetric partition function on $S^2\times S^1$  with periodic boundary conditions along the $S^1$ direction. Equivalently, it is a trace over the Hilbert space on $S^2$ in radial quantization~\cite{Bhattacharya:2008zy}:
\begin{align}
  Z=\operatorname{Tr}_{S^2}\Bigl[(-1)^F e^{-\beta'\{\mathcal{Q},\mathcal{Q}^\dagger\}}e^{-\beta(\epsilon+j_3)}t_I^{F_I}\Bigr]\,,
\end{align}
where $F$ is the fermion number. The supercharge $\mathcal{Q}$ satisfies the anticommutation relation
\begin{align}
  \{\mathcal{Q},\mathcal{Q}^\dagger\}=\epsilon-h_3-j_3\,.
\end{align}
Here, $\epsilon$ is the energy in radial quantization, $h_3$ is the $SO_\mathcal{R}(6)$ R-charge, and $j_3$ is the third component of the angular momentum on $S^2$. The $F_I$'s denote the combinations of global charges that commute with the chosen $\mathcal{Q}$ and its conjugate $\mathcal{S}\equiv\mathcal{Q}^\dagger$, and $t_I$'s are the associated fugacities.

Using the technique of the supersymmetric localization on the Coulomb branch, the ABJM theory's SCI reduces to the following matrix integral~\cite{Bhattacharya:2008bja, Kim:2009wb}
\begin{align},
  Z=&\frac{1}{(N!)^2}\sum_{\mathfrak{m}_i,\tilde{\mathfrak{m}}_i\in\mathbb{Z}^N}\oint\left(\prod_{i=1}^N\frac{ds_i}{2\pi is_i}\frac{d\tilde{s}_i}{2\pi i\tilde{s}_i}s_i^{k\mathfrak{m}_i}\tilde{s}_i^{-k\tilde{\mathfrak{m}}_i}\right)\times\nonumber\\
  &\qquad\times\prod_{i\neq j}q^{-\frac{1}{2}|\mathfrak{m}_{i}-\mathfrak{m}_{j}|-\frac{1}{2}|\tilde{\mathfrak{m}}_{i}-\tilde{\mathfrak{m}}_{j}|}\Bigl(1-s_is_j^{-1}q^{|\mathfrak{m}_{i}-\mathfrak{m}_{j}|}\Bigr)\Bigl(1-\tilde{s}_i\tilde{s}_j^{-1}q^{|\tilde{\mathfrak{m}}_{i}-\tilde{\mathfrak{m}}_{j}|}\Bigr)\times\nonumber\\
  &\qquad\times\prod_{i,j=1}^Nq^{\frac{1}{4}|\mathfrak{m}_i-\tilde{\mathfrak{m}}_j|+\frac{1}{4}|-\mathfrak{m}_i+\tilde{\mathfrak{m}}_j|}\frac{\bigl(s_i^{-1}\tilde{s}_jt_{1,2}^{-1}q^{2+|\mathfrak{m}_i-\tilde{\mathfrak{m}}_j|};q^2\bigr)}{\bigl(s_i\tilde{s}_j^{-1}t_{1,2}q^{|\mathfrak{m}_a-\tilde{\mathfrak{m}}_b|};q^2\bigr)}\frac{\bigl(s_i\tilde{s}_j^{-1}t_{3,4}^{-1}q^{2+|-\mathfrak{m}_i+\tilde{\mathfrak{m}}_j|};q^2\bigr)}{\bigl(s_i^{-1}\tilde{s}_jt_{3,4}q^{|-\mathfrak{m}_i+\tilde{\mathfrak{m}}_j|};q^2\bigr)}\,,\label{eq:MatrixIntegral}
\end{align}
where $q=e^{-\beta}$, $t_I=e^{\Delta_I}$, $s_i=e^{iy_i}$ $\tilde{s}_i=e^{i\tilde{y}_i}$. The integers $\mathfrak{m}_i$ and $\tilde{\mathfrak{m}}_i$ are magnetic fluxes of Dirac momopoles. $(a;z)\equiv\prod_{n=0}^\infty(1-az^n)$ denotes the $q$-Pochhammer symbol. The second line collects the contributions from the vector multiplets, while the last line originates from the chiral multiplets. For simplicity, we consider the case of equal chemical potentials. 
The BPS condition 
\begin{align}
  \sum_{I=1}^{4}\Delta_I+2\beta=2\pi i 
\end{align}
then reduces to
\begin{align}
  \Delta\equiv\Delta_I=\frac{\pi i-\beta}{2}\label{eq:BPS condition }
\end{align}
for every $I$.

Motivated by the construction of multi-cut saddles in~\cite{Choi:2021rxi}, we introduce shifted eigenvalue ansatz
\begin{align}
  y_i\to y_i+\frac{r}{K}\,,\quad \tilde{y}_i\to \tilde{y}_i+\frac{r}{K}\,,\quad r=0,\cdots,K-1\,.
\end{align}
This shift leaves the matrix integral \eqref{eq:MatrixIntegral} in the large-$N$ limit invariant, because the holonomies of the $U(N)$ gauge group always appear together with their inverses. For the SCI of the 4d $\mathcal{N}=4$ super-Yang-Mills (SYM) theory, the multi-cut saddle points in the large-$N$ limit with equal angular momenta are known to be in one-to-one correspondence with the Bethe roots of the matrix integral and preserve the $\mathbb{Z}^{(1)}_K$ 1-form center symmetry~\cite{Choi:2021rxi}. In the case of the ABJM SCI, however, a complete Bethe ansatz formulation has not yet been established. The shifted eigenvalue ansatz introduced here, therefore, offers a constructive hint toward deriving the corresponding Bethe ansatz equations for the ABJM SCI. The corresponding rescaled chemical potentials are defined as
\begin{align}
  \Delta_K \equiv K\left(\Delta-2\pi i\frac{K-1}{4K}\right)\,,\quad \beta_K \equiv K\beta\,,\label{eq:multi-cut chemical potentials}
\end{align}
which again satisfy the BPS condition  $2\Delta_K+\beta_K=\pi i$.

In the large-$N$ limit, the $K=1$ saddle of the matrix integral corresponds to the AdS$_4$ BPS black hole solution described in~\cite{Chong:2004na}. The saddles with $K>1$, on the other hand,  describe configurations in which the eigenvalues $y_i$ and $\tilde{y}_i$ condense into multiple distinct cuts or clusters, each representing a densely packed branch of distributed eigenvalues.

Following the treatment of~\cite{GonzalezLezcano:2022hcf}, the large-$N$ matrix integral can be evaluated in the Cardy-like limit $\beta\to0$ through a saddle-point analysis. More precisely, we introduce the  variables
\begin{align}
  \mathfrak{m}_i,\tilde{\mathfrak{m}}_i\to\frac{N^\alpha}{\beta}x_i\,,
\end{align}
and the replacement
\begin{align}
  \sum_{i=1}^N\to\int_{-\infty}^\infty dx\rho(x)\,,
\end{align}
where $\rho(x)$ is the normalized density of the continuous variables $x$. After the same appropriate manipulations as in~\cite{GonzalezLezcano:2022hcf}, the index can be reduced to
\begin{align}
  Z=\int \mathcal{D}\rho\mathcal{D}(\delta y)\,\exp\Bigl(-S_{\text{eff}}[x;\rho(x),\delta y(x)]\Bigr)\,,
\end{align}
where the effective action is organized as an expansion in $N$ at the leading order in $\beta$
\begin{align}
  S_{\text{eff}}[x;\rho(x), y(x)]=N^{3/2}\int dx \mathcal{L}^{(3/2)}[x;\rho(x),\delta y(x)]+\mathcal{O}(N^{3/2})\,.
\end{align}
More explicitly, 
\begin{align}
  \mathcal{L}^{(3/2)}=\frac{i}{\beta}k\,\rho(\mu-x\delta y)+\frac{i}{\beta}\rho^2\left(\pi (\delta y)^2-2i\Delta^3\right)-(\delta y)^2\rho^2-\frac{2i}{3}\pi\beta\rho^2\,,
\end{align}
where $\delta y=\tilde{y}(x)-y(x)$. Here, we have used a Lagrange multiplier to insert the normalization condition for the density of states $\rho(x)$ into the effective action as
\begin{align}
  S_\mu=N^{3/2}\frac{ik}{\beta}\mu\Bigl(\int dx\rho(x)-1\Bigr)\,.
\end{align}
Within this multi-cut framework, we further define $\omega \equiv -2\beta$, and then the Lagrangian becomes
\begin{align}
  \mathcal{L}^{(3/2)}=-\frac{2i}{\omega_K}k\,\rho(\mu-x\delta y)-\frac{2i}{\omega_K}\rho^2\left(\pi (\delta y)^2-2i\Delta_K^3\right)-(\delta y)^2\rho^2+\frac{i}{3}\pi\omega_K\rho^2\,,
\end{align}
with $\omega_K\equiv-2\beta_K$.

The saddle-point equations derived from this effective action admit three distinct types of solutions, depending on the value of $x$. 
The ``inner solution'' reads
\begin{align}
  \rho^{(\text{i})}(x)=\frac{3k(2\pi-i\omega_K)\mu}{\pi(2\pi-i\omega_K)\omega_K^2+48\Delta_K^4}\,,\quad \delta y^{(\text{i})}(x)=\frac{\pi x\omega_K^2+12ix\Delta_K^3}{3(2\pi-i\omega_K)\mu}\,.
\end{align} 
The ``left-tail solution'' and the ``right-tail solution'' are respectively 
\begin{align}
  \delta y^{(\text{l})}(x)=i\Delta_K+\mathcal{O}(e^{-N})\,,\quad \rho^{(\text{l})}(x)=\frac{3k(\mu-ix\Delta_K)}{\pi\omega_K^2}\,,
\end{align}
\begin{align}
  \delta y^{(\text{r})}(x)=-i\Delta_K+\mathcal{O}(e^{-N})\,,\quad \rho^{(\text{r})}(x)=\frac{3k(\mu+ix\Delta_K)}{\pi\omega_K^2}\,.
\end{align}
Substituting all these solutions in different segments into the effective action and imposing the normalization condition to fix $\mu$, we obtain the large-$N$ free energy in the Cardy-like limit
\begin{align}
  \log\,Z=-N^{3/2}\frac{ik^2\mu^3}{3\,\omega_K\Delta_K^4}+\mathcal{O}(\omega_K^0)=N^{3/2}\frac{2i\sqrt{2k}\Delta_K^2}{3\,\omega_K}+\mathcal{O}(\omega_K^0)\,.
\end{align}
  Up to an overall factor of $-2$, this expression agrees with the result from the entropy function~\cite{Choi:2018fdc}
  \begin{align}
    \log Z=-iN^{3/2}k^{1/2}\frac{4\sqrt{2}\Delta^2}{3\, \omega}\,.
  \end{align}
For a given $K$-cut saddle, we further shift $\omega\to\omega-2\pi ir/K$, with $r=0,\cdots,K-1$, which yields the free energy labeled by $(K,r)$
\begin{align}
  \log\,Z(\omega)=-N^{3/2}i\sqrt{2k}K\frac{\left(\omega-\frac{2\pi i r}{K}+\frac{2\pi i}{K}\right)^2}{12\left(\omega-\frac{2\pi i r}{K}\right)}\,.
\end{align}
To extract the degeneracy of states, we perform a Legendre transformation to the entropy function
\begin{align}
  S(\omega)=\log Z-\Bigl(\sum_I\Delta_IQ_I+\omega J\Bigr)\,.
\end{align}
With equal chemical potentials, the terms in parentheses simplify to
\begin{align}
  \sum_I  \Delta_I Q_I + \omega J=\frac{\omega}{4}j+2\pi iQ\,,
\end{align}
where we have defined $j\equiv 4(Q+J)$.
Consequently, the density of states is given by an inverse Laplace transform
\begin{align}
  \Omega(\tilde{j})=\int\frac{d\omega}{8\pi i}Z(\omega)\,e^{\frac{\omega-2\pi i}{4}\tilde{j}}\,.
\end{align}
Finally, the microcanonical SFF in this setting takes the form
\begin{align}
  Y_{j,\Delta}(\tau)=\int_\mathcal{C}d\omega\frac{\Delta}{4\sqrt{2\pi}i}e^{\frac{\omega-i\tau}{4}j+\frac{(\omega-i\tau)^2}{32}\Delta^2}\,,\qquad\tau \equiv 4t+2\pi\,.
\end{align}
Evaluating the integral via a saddle-point approximation leads to the compact expression
\begin{align}
  Y_{j,\Delta}(\tau)\sim\sum_{K\in\mathbb{N},r}\exp\left[\frac{\omega_*}{4}j+\frac{\omega_*^2}{32}\Delta^2+\log Z(\omega_*+i\tau)\right]\,,
\end{align}
where the sum runs over all multi-cut saddles labeled by $(K,r)$, and $\omega_*$ denotes the saddle-point value of $\omega$ determined by the extremization condition.

\begin{figure}[htb!]
    \begin{center}
      \includegraphics[width=0.85\textwidth]{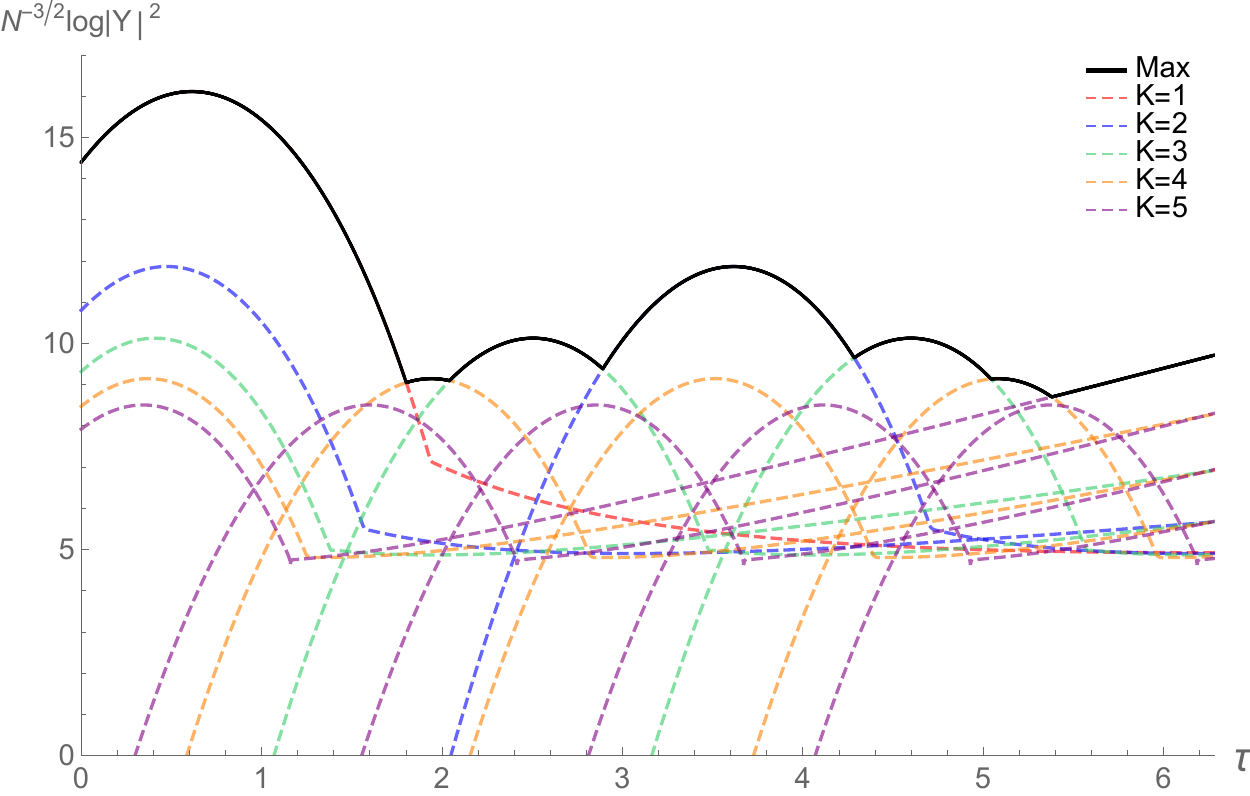}
    \end{center}
    \caption{\label{fig:SFF-ABJM}  Spectral form factor as a function of time plotted from $K$-saddles of the ABJM superconformal index with $K=1,\cdots,5$ and  $\tau\in(0,2\pi)$. We take $k=1$, $N=1000$, $j=10N^{3/2}$, and $\Delta=10N^{3/4}$.}
    \end{figure}

 Fig.~\ref{fig:SFF-ABJM} illustrates the characteristic behavior of the spectral form factor normalized by $N^{-3/2}$ with equal chemical potentials. After an initial transient rise, the SFF exhibits the expected rapid decay at early times, which is a regime commonly termed the ``slope''. As time evolves,  this early-time behavior is superseded by contributions from the  $K>1$ saddles. 
  These multi-cut saddles dominate the SFF during intermediate times, thereby preventing it from decaying further. Holographically, these multi-cut saddles are expected to be  dual to $\mathbb{Z}_K$-orbifolded Euclidean black holes in AdS$_4$, which obey the same boundary conditions as the BPS black hole~\cite{Chong:2004na}. A qualitatively similar mechanism, where subdominant saddles emerge to stabilize the SFF, has been observed in both free Yang-Mills theory \cite{Chen:2022hbi} and SYM theory \cite{Choi:2022asl} in the large-$N$ limit.

\begin{figure}[htb!]
    \begin{center}
      \includegraphics[width=0.85\textwidth]{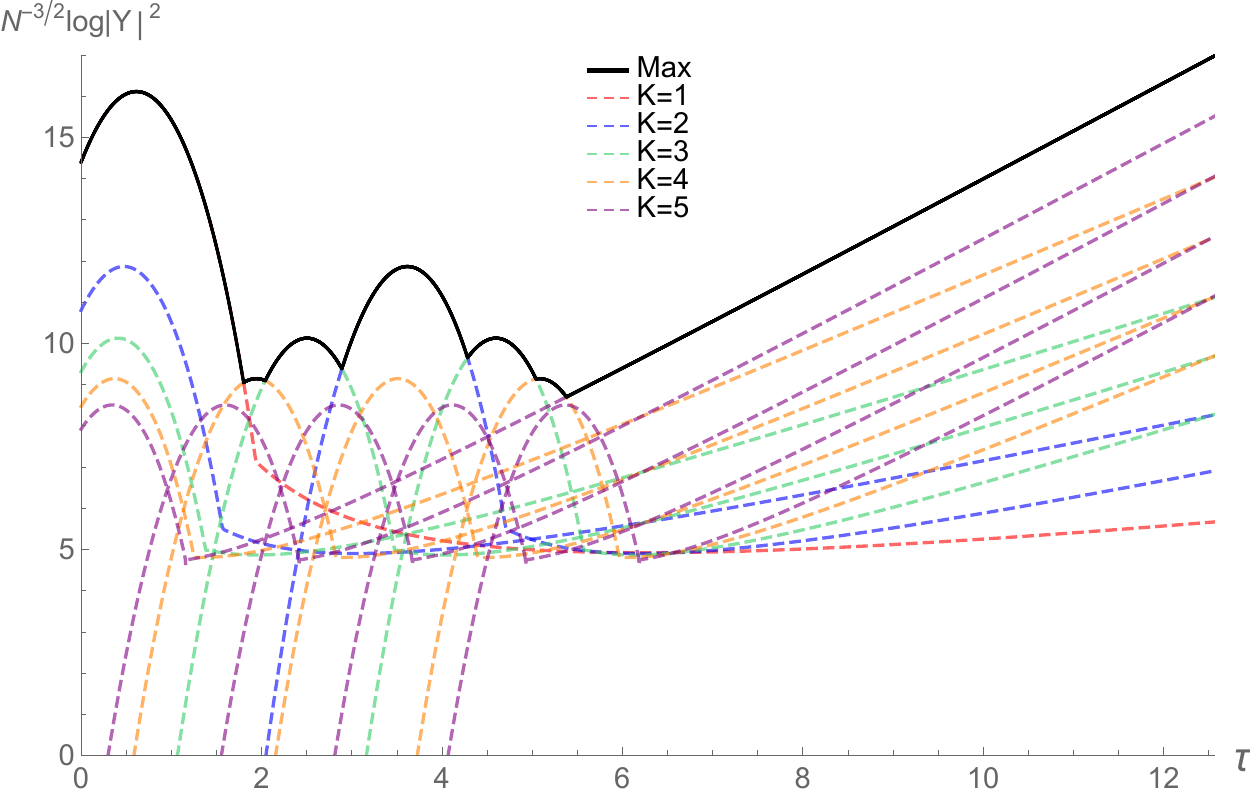}
    \end{center}
    \caption{\label{fig:SFF-ABJM large t} Spectral form factor as a function of time plotted from $K$-saddles of the ABJM superconformal index with $K=1,\cdots,5$ and  $\tau\in(0,4\pi)$. We take $k=1$, $N=1000$, $j=10N^{3/2}$, and $\Delta=10N^{3/4}$.}
    \end{figure}

We also observe that the late-time behavior of the SFF in the ABJM model exhibits a sharp increase, which we refer to as the ``ramp'', as shown in Fig.~\ref{fig:SFF-ABJM large t}. At these late stages, saddles with progressively larger values of $K$ become increasingly dominant.
The SFF does not saturate to a constant ``plateau'' within the time scales accessible to our analysis. This absence of a plateau is because the plateau time scales as $t_p\sim e^{N^{3/2}/(\beta^2)}$, which grows exponentially in $N^{3/2}$ and inversely in $\beta^2$. As a result, the ramp persists over a parametrically extended duration, far exceeding the time windows typically accessible through direct numerical or analytic methods in the large-$N$ limit. This extended ramp behavior highlights a key difference in the late-time spectral correlations of the ABJM model compared to systems that exhibit a clear plateau within accessible time scales.

\section{Discussion}\label{sec:discussion}

In this manuscript, we have investigated the microcanonical spectral form factor (SFF) in the context of the ABJM theory in the large-$N$ limit. Building upon the framework of the superconformal index in the Cardy-like limit, we have systematically incorporated multi-cut saddle-point configurations and derived their contributions to the SFF. Our analysis reveals a rich dynamical structure:

\begin{itemize}
  \item The SFF exhibits the characteristic early-time slope, followed by a regime where contributions from $K>1$ multi-cut saddles, which are holographically dual to $\mathbb{Z}_K$-orbifolded Euclidean black holes in AdS$_4$, become dominant, thereby preventing further decay.
  
  \item The ABJM model exhibits pronounced late-time growth of the SFF, driven by the increasing dominance of saddles with larger $K$.
  
  \item These results highlight the role of multi-cut saddles in governing the fine-grained spectral dynamics and provide a concrete setting in which the late-time behavior of the SFF encodes information beyond that captured by random-matrix universality.
\end{itemize}

There are several interesting open problems that we defer to future studies. First, a first-principle field-theoretic understanding of the spectral correlations responsible for the late-time rise of the SFF in ABJM remains to be developed. The superconformal index considered here only counts 1/12-BPS states; we expect that including more states should be responsible for the emergent plateau behavior. Moreover, although the $K>1$ saddles are expected to be dual to orbifolded black holes, a detailed bulk interpretation remains lacking. Do they correspond to a dynamical network of wormholes or to resonant geometries in the Euclidean path integral? Finally, similar studies on SFF can be applied to other AdS$_4$ black holes, including the magnetically charged STU black holes \cite{Benini:2015eyy} and non-extremal black holes at high temperatures \cite{Nian:2025iei}.

\acknowledgments

The author thanks Guang-Shang Chen, Ping Gao, Cong-Yuan Yue, and Yu-Xuan Zhang for useful discussions.  Special thanks to Jun Nian for his instructions throughout this work. This work is supported in part by the NSFC under grants No.~12375067, No.~12147103, and No.~12247103.



\bibliographystyle{utphys}
\bibliography{ref}



\end{document}